# An automatic deep learning-based workflow for glioblastoma survival prediction using pre-operative multimodal MR images

Jie Fu[1*], Kamal Singhrao[1], Xinran Zhong[2], Yu Gao[2], Sharon Qi[1], Yingli Yang[1], Dan Ruan[1], John H Lewis[3*]

1 Department of Radiation Oncology, University of California, Los Angeles, Los Angeles, CA, USA, 90095

2 University of California, Los Angeles, Department of Radiological Sciences, Los Angeles, CA, USA, 90095

3 Department of Radiation Oncology, Cedars Sinai Medical Center, Los Angeles, CA, USA, 90048

*Email: jiefu@mednet.ucla.edu and john.h.lewis@cshs.org

## Abstract

**Purpose:** Radiomic features have demonstrated promise in improving tumor classification, treatment outcome prediction, and survival analysis. However, most existing radiomic studies were based on features extracted using manual contours. Large inter-observer segmentation variations may lead to inconsistent features and introduce additional challenges in constructing robust prediction models. Here, we proposed a fully automatic workflow for glioblastoma (GBM) survival prediction using deep learning (DL) methods.

**Methods:** 285 glioma (210 GBM, 75 low-grade glioma) patients were included. 163 of the GBM patients had overall survival (OS) data. Every patient had four pre-operative MR scans and manually drawn tumor contours. For automatic tumor segmentation, a 3D convolutional neural network (CNN) was trained and validated using 122 glioma patients. The trained model was applied to the remaining 163 GBM patients to generate tumor contours. The handcrafted and DL-based radiomic features were extracted from auto-contours using explicitly designed algorithms and a pre-trained CNN respectively. 163 GBM patients were randomly split into training (n=122) and testing (n=41) sets for survival analysis. Cox regression models with regularization techniques were trained to construct the handcrafted and DL-based signatures. The prognostic power of the two signatures was evaluated and compared.

**Results**: The 3D CNN achieved an average Dice coefficient of 0.85 across 163 GBM patients for tumor segmentation. The handcrafted signature achieved a C-index of 0.64 (95% CI: 0.55-0.73), while the DL-based signature achieved a C-index of 0.67 (95% CI: 0.57-0.77). Unlike the handcrafted signature, the DL-based signature successfully stratified testing patients into two prognostically distinct groups (*p*-value<0.01, HR=2.80, 95% CI: 1.26-6.24).

**Conclusions:** The proposed 3D CNN generated accurate GBM tumor contours from four MR images. The DL-based signature resulted in better GBM survival prediction, in terms of higher C-index and significant patient stratification, than the handcrafted signature. The proposed automatic radiomic workflow demonstrated the potential of improving patient stratification and survival prediction in GBM patients.

## 1. Introduction

Glioma is the most common type of primary brain tumor in adults. It arises from glial cells, normally astrocytes and oligodendrocytes. According to the World Health Organization guideline[1], glioma can be classified into grade I to IV based on the histological characteristics. Glioblastoma multiforme (GBM) is the most aggressive, grade IV, glioma. It accounts for 81% of malignant brain tumors[2]. Despite extensive efforts, GBM patient prognoses remain dismal. The median overall survival (OS) is 14 to 16 months after diagnosis[3]. The 5-year survival rate is below 5%[4]. It is beneficial to build survival prediction models for assisting therapeutic decisions and disease management in GBM patients.

Magnetic resonance imaging (MRI) is the preferred imaging modality for GBM diagnosis and monitoring. Radiomic features extracted from MR images using advanced mathematical algorithms may uncover tumor characteristics that fail to be appreciated by the naked eye. Recent studies demonstrated that these features can assist in tumor grading and predict the isocitrate dehydrogenase (IDH) genotype for glioma patients[5,6]. Many studies have investigated the association of radiomic features with the survival outcomes of GBM patients. Most of them were based on radiomic features extracted using explicitly designed, or "handcrafted", algorithms[7,8]. These handcrafted features are normally low-level image features that are limited to current human knowledge.

Another type of radiomic feature is extracted using deep learning (DL) methods like convolutional neural networks (CNNs)[9, 10]. These high-level features may have higher prognostic power than the handcrafted features. A study demonstrated that DL-based radiomic features extracted via transfer learning achieved promising performance in GBM survival prediction[11]. However, the radiomic features were extracted from manual tumor contours in this study. Manual tumor segmentation is not only time consuming but also sensitive to intra-observer and inter-observer variabilities. Studies have shown that inter-observer contour variation could result in many inconsistent radiomic features[12, 13]. This may introduce additional challenges in constructing robust prediction models. Developing an automatic GBM tumor segmentation model could erase this concern.

In this study, we proposed an automatic workflow for achieving GBM survival prediction using four pre-operative MR images. A 3D CNN was proposed and trained for automatically generating GBM tumor contours. We investigated whether radiomic features extracted using the auto-contours were associated with GBM OS. The handcrafted and DL-based radiomic features were extracted from the auto-contours and used to construct two separate Cox regression models. The prognostic power of the constructed signatures was evaluated and compared for better model performance.

## 2. Materials and methods

2.1. Dataset

285 glioma patients were acquired from the BRaTS2018 challenge[14–16]. 210 patients had GBM, and the remaining 75 patients had low-grade (grade II-III) glioma (LGG). Each patient had four MR images acquired. These included T1-weighted (T1w), contrast-enhanced T1-weighted (CE-T1w), T2-weighted (T2w), and fluid-attenuated inversion recovery (FLAIR) MR images. Patient images were acquired with different clinical protocols and various scanners from multiple institutions. For each patient, MR images were co-registered, resampled to 1 mm$^3$ resolution, and skull-stripped. The final image dimension was 240×240×155. All patients have three ground truth tumor subregion labels (edema, enhancing tumor core, and non-enhancing tumor core) approved by experienced neuro-radiologists. OS data was available for 163 GBM patients.

We applied N4 bias correction[17] on all images, except the FLAIR images, to remove low-frequency inhomogeneity. Each MR image was normalized to have zero mean and unit standard deviation in the brain voxels. Figure 1 shows the transverse slice of four preprocessed MR images and the corresponding tumor labels.

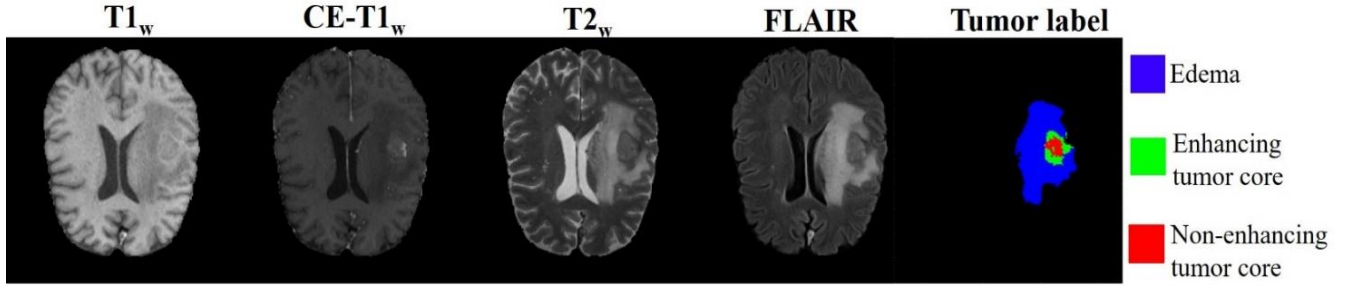

Figure 1. Transverse slices of preprocessed T1w, CE-T1w, T2w, FLAIR images along with the corresponding ground truth labels for edema, enhancing tumor core, and non-enhancing tumor core for a representative case.

2.2. 3D CNN for tumor segmentation

Figure 2 shows the architecture of the 3D CNN proposed for tumor segmentation. It contains 27 convolutional layers, forming an encoder and decoder architecture. Instance normalization layers[18] and residual shortcuts[19] are implemented to improve model performance. The proposed 3D CNN can be trained to perform an end-to-end mapping, converting the concatenation of four preprocessed images to four probability maps for three tumor subregion labels and background labels.

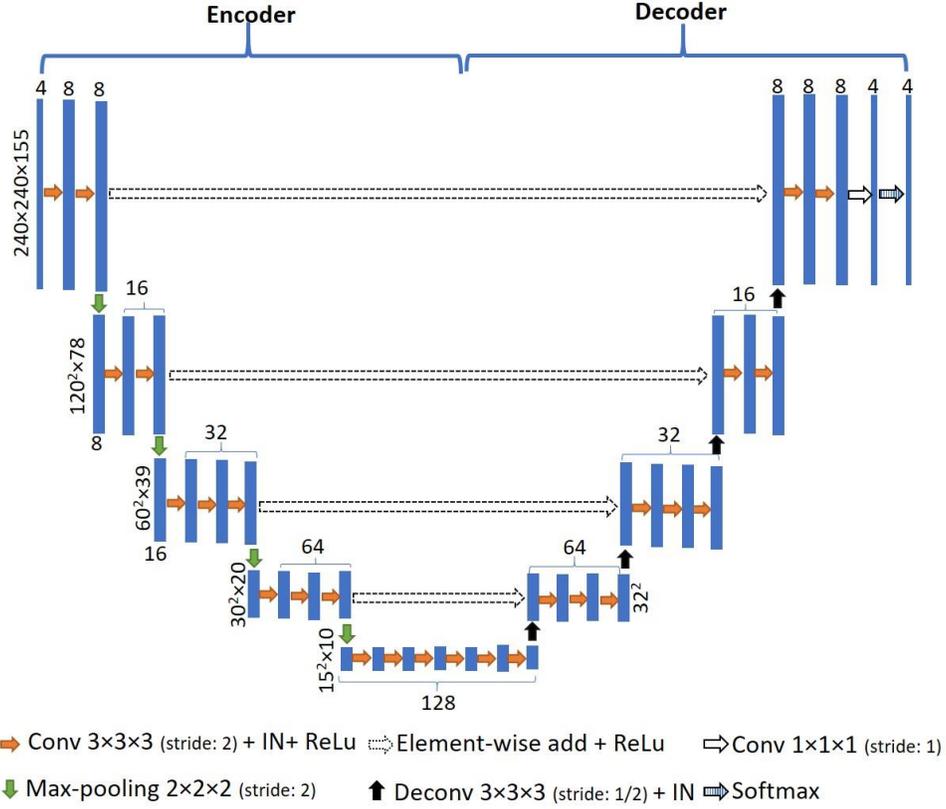

Figure 2. The overall 3D CNN architecture. Four MR images are concatenated and input into the 3D CNN containing 27 convolutional layers. The model generates four probability maps. Each filled box represents a set of 4D feature maps, the numbers and dimensions of which are shown. The window size and the stride for convolutional, maxpooling, and deconvolutional layers are also presented. Conv, convolutional layer; IN, instance normalization layer; ReLu, rectified linear unit; Maxpool, maxpooling layer; Deconv, deconvolutional layer.

In the model training stage, 122 patients without OS data were randomly split into a training set of 105 patients (75 LGG patients and 30 GBM patients) and a validation set of 17 GBM patients. The validation set was used for tuning hyperparameters including the stopping epoch number and the learning rate. The Adam stochastic gradient descent method[20] was used to minimize the multi-Dice loss,

$$\text{loss} = \frac{1}{4}\sum_{i=1}^{4}(1 - \frac{2\sum_{j=1}^{N} P_{ij} \times L_{ij}}{\sum_{j=1}^{N} P_{ij} + \sum_{j=1}^{N} L_{ij}}),$$

(1)

where $P_{ij}$ is the probability, after Softmax layers, of the voxel j being the label i; $L_{ij}$ is the ground truth label, 0 or 1, of the voxel j being the label i; N is the voxel number.

The 3D CNN was trained using the training set with a batch size of 1 for 50 epochs. These hyperparameters were determined using the validation set. The trained network was applied to the remaining 163 GBM patients (all of which have corresponding OS data) to generate their tumor subregion labels. The tumor contour was acquired by merging the three subregion labels. Model accuracy was evaluated using the Dice coefficient,

$$\text{Dice} = \frac{2(V_{ground} \cap V_{pred})}{V_{ground} + V_{pred}},$$

(2)

where $V_{ground}$ and $V_{pred}$ are the ground truth tumor contour and 3D CNN predicted tumor contour, respectively.

2.3. Radiomic feature extraction

2.3.1. Handcrafted features

1106 handcrafted features were extracted from four MR images using the PyRadiomics[21] package (version 2.1.2) for all 163 GBM patients. These features were extracted from the 3D CNN-predicted tumor contour and contained 14 shape-based features, 72 first-order statistical features, 292 second-order statistical (textural) features, and 728 high-order statistical features. Shape-based features represented the shape characteristics of the tumor contour. First-order statistical features represented the characteristics of the tumor intensity distribution. Textural features were extracted based on gray level co-occurrence, gray level size zone, gray level run length, gray level dependence, and neighborhood gray-tone difference matrices. They represented the characteristics of the spatial intensity distributions. High-order statistic features were extracted from the images filtered using Laplacian of Gaussian (LoG) filters.

2.3.2. DL-based features

1472 DL-based features were extracted using a pre-trained classification CNN, VGG19[22], for all 163 GBM patients in the testing set. Figure 3 shows the model architecture and feature extraction scheme. VGG19 contains 16 convolutional layers and 3 fully-connected layers. 5 max-pooling layers are used to achieve partial translational invariance, reduce model memory usage, and prevent overfitting. For each patient, we selected a square region of interest (ROI) from the transverse slice that had the largest 3D CNN-predicted tumor contour area. The dimension of the ROI was set as the maximum dimension of the tumor contour on the selected slice. We then extrapolated the ROIs of FLAIR, T2w, and CE-T1w MR images to 224×224, mapped the pixel intensity to the range [0, 255], and concatenated them. The concatenation was input into the pre-trained VGG19 for feature extraction. As shown in Figure 3, DL-based features were extracted by average-pooling the 5 feature maps after max-pooling layers. Each feature map generated a vector after average-pooling. Five feature vectors were first normalized with their Euclidean norms and then concatenated to form a single feature vector. DL-based features were acquired by normalizing the single feature vector with its Euclidean norm.

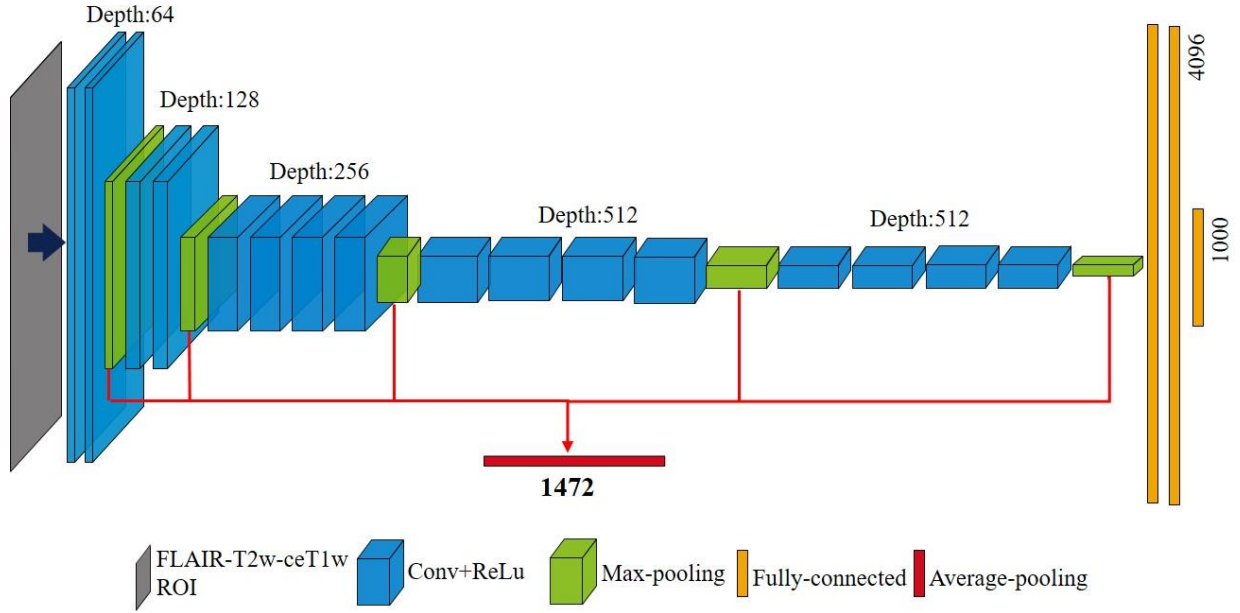

Figure 3. DL-based feature extraction scheme using VGG19. VGG19 contains 16 convolutional layers, 5 max-pooling layers, and 3 fully-connected layers. The average-pooling layers were used for extracting DL-based features. Feature maps and feature vectors after every layer are shown as cuboids and rectangles, respectively. The feature map depth and feature number are shown. A concatenation of FLAIR, T2w, and CE-T1w ROIs was input into the pre-trained VGG19 for feature extraction. 1472 DL-based features were extracted from max-pooling feature maps by average-pooling along the spatial dimensions. Conv, convolutional layer; ReLu, rectified linear unit.

## 2.4. Survival prediction model

The 163 GBM patients with available OS data were randomly split into a training set of 122 patients and a testing set of 41 patients. Each feature was normalized using the mean and standard deviation of the training set. Since a large number of features may lead to overfitting, we pre-selected a subset of features having the highest univariate C-index. Higher C-index values indicate features with higher prognostic power. The Cox regression model with regularization was trained using the selected features to construct a radiomic signature for survival prediction in GBM patients. The radiomic signature is a linear combination of the features weighted by the Cox regression model coefficients. We tested three regularization techniques: Ridge, Elastic Net, and Least Absolute Shrinkage and Selection Operator (LASSO). The number of the pre-selected features, the regularization technique, and the corresponding regularization parameters were chosen with 5-fold cross-validation using the training set. Two Cox regression models were trained using either handcrafted features or DL-based features. The resulting radiomic signatures are referred to as the handcrafted signature and the DL-based signature, respectively. The prognostic power of the two constructed radiomic signatures was compared using the C-index and the average areas under the receiver operating curves (AUCs) at different survival time points on the testing set. Paired t-tests and DeLong tests were conducted to test the significance of the differences in C-index and receiver operating curves, respectively. A threshold on the radiomic signature can be set using the training set for patient stratification. We investigated two thresholds: one selected using the X-tile software[23], and the other defined by the median signature value of the training patients. The X-tile software selected the optimal threshold by selecting the highest $X^2$ value of the data divisions. The chosen thresholds were then used to stratify the testing patients into high-risk and low-risk groups. Log-rank tests were conducted to test the difference between the two risk groups for significance.

## 3. Results

3.1. OS statistics

The median and mean of OS were 367.0 days and 416.5 days in the training set, and 362.0 days and 442.1 days in the testing set, respectively. A Mann-Whitney U test indicated that there was no significant difference in OS between two datasets (*p*-value=0.83).

3.2. Tumor segmentation

Figure 4 compares the ground truth segmentation and 3D CNN predicted tumor contour for three patients in the testing set. Contours generated by the trained 3D CNN were smoother than the corresponding ground truth contours. They had similar shapes based on visual inspection.

The Dice coefficients of tumor contours for the training, validation, and testing sets are summarized in Table 1. The predicted contours achieved the Dice coefficient of 0.85±0.09 for 163 GBM patients in the testing set.

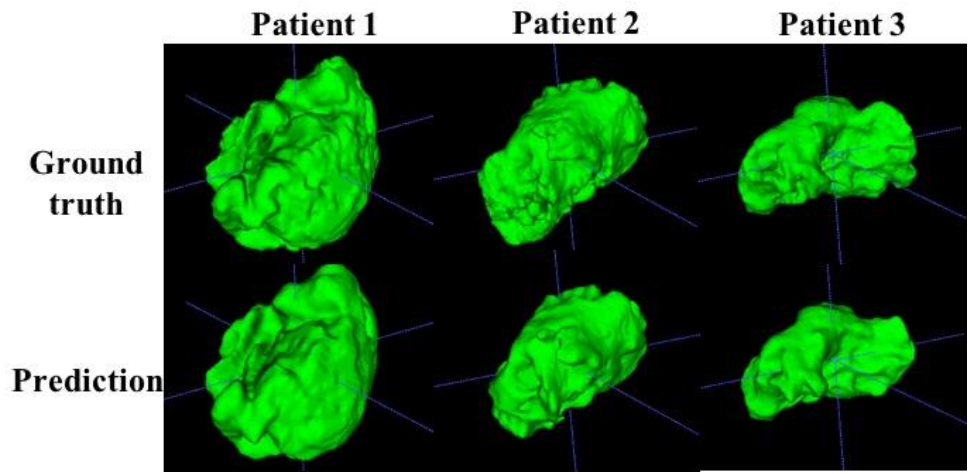

Figure 4. Ground truth contour (top) and 3D predicted contour (bottom) for 3 GBM patients

|  | Training | Validation | Testing |
|---|---|---|---|
|  | (75 LGG and 30 GBM) | (17 GBM) | (163 GBM) |
| **Dice** | 0.91±0.03 | 0.90±0.06 | 0.85±0.09 |

Table 1. Dice coefficients of the tumor contour for the training, validation, and testing sets. Results were averaged and showed in (mean ± SD) format.

3.3. Survival prediction

The handcrafted signature achieved a C-index of 0.64 (95% confidence intervals [CI]: 0.55-0.73) on the testing set, while the DL-based signature achieved a C-index of 0.67 (95% CI: 0.57-0.77). A paired t-test showed that there was no significant difference in these C-indexes ($p$-value=0.27).

Table 2 shows the AUCs of the signatures evaluated at three survival time points (183, 356, and 548 days) on the testing set. The handcrafted signature achieved higher AUCs evaluated at the OS of 183 days and 365 days, while the DL-based signature achieved higher AUC evaluated at 548 days. $p$-values of DeLong tests indicated that there was no significant difference in all AUC results.

| **Relapse time (days)** | **Handcrafted** | **DL-based** | ***p*-value** |
|---|---|---|---|
| **183** | 0.76 (0.60-0.92) | 0.75 (0.57-0.93) | 0.93 |
| **365** | 0.77 (0.62-0.92) | 0.74 (0.58-0.90) | 0.72 |
| **548** | 0.58 (0.37-0.79) | 0.68 (0.50-0.86) | 0.26 |

Table 2. AUC results, evaluated at three survival time points, for the handcrafted and DL-based signatures. 95% confidence intervals are shown in the parentheses. $p$-values of the corresponding DeLong tests are shown in the last column.

The testing patients were split into high-risk and low-risk groups based on signature thresholds. Figure 5 shows the Kaplan-Meier survival curves of the two risk groups. There was no significant association between the risk groups, stratified by thresholding the handcrafted signature, and the patient OS. (X-tile: $p$-value=0.31, hazard ratio [HR]=1.44, 95% CI: 0.71-2.91; Median: p-value=0.20, HR=1.51, 95% CI: 0.80-2.87). On the other hand, thresholds on the DL-based signature resulted in significant stratification of patients into two prognostically distinct groups (X-tile: $p$-value<0.01, HR=2.80, 95% CI: 1.26-6.24; Median: $p$-value=0.02, HR=2.16, 95% CI: 1.12-4.17).

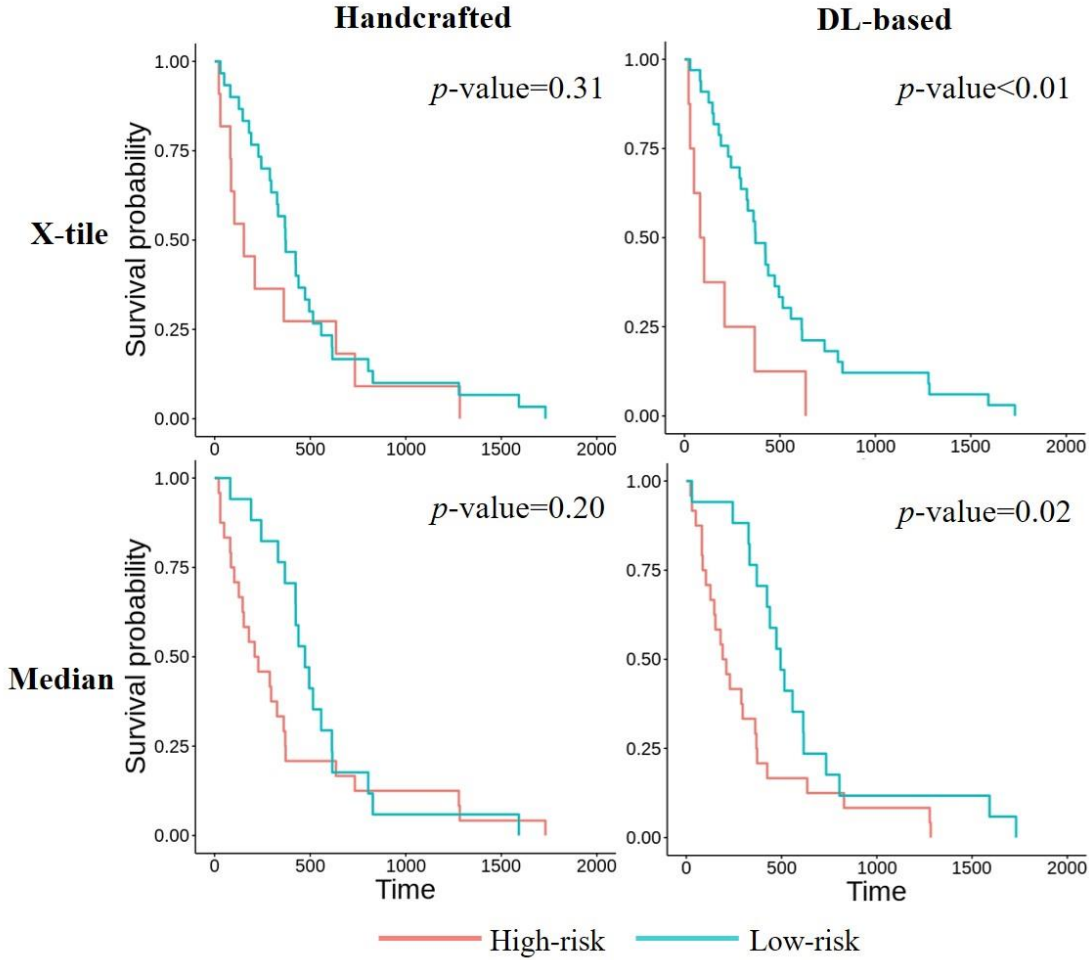

Figure 5. Kaplan-Meier survival curves of the testing patients. Patients were stratified into two risk groups based on thresholds of the handcrafted signature or the DL-based signature. The top row shows the stratification based on the threshold generated by X-tile software, and the bottom row shows the stratification based on the median signature value. *p*-values of the corresponding log-rank tests are shown.

## 4. Discussion

In this paper, we investigated deep learning approaches for achieving automatic GBM survival prediction. A 3D CNN was proposed and trained using 105 glioma patients for generating GBM tumor contours from four pre-operative MR images. The trained model was applied to 163 GBM patients to generate their tumor contours for survival analysis. The proposed 3D CNN achieved accurate GBM tumor contours, with a mean Dice coefficient of 0.85 for the 163 GBM patients. We extracted handcrafted and DL-based radiomic features from the MR images using the auto-contours for these patients. Two Cox regression models were trained using the extracted features to construct the handcrafted and DL-based signatures for survival prediction. The DL-based signature achieved a C-index of 0.67 and successfully stratified the GBM patients into high-risk and low-risk groups.

We included 75 LGG patients for training the 3D segmentation CNN as our test showed that the model trained with both 75 LGG patients and 30 GBM patients achieved better performance than the model trained with 30 GBM patients alone. This is expected since LGG and GBM have a similar appearance in MR images. Our 3D CNN could generate three tumor subregion labels. However, the accuracy of segmenting subregion labels using the 3D CNN was low, with the mean Dice coefficient of the tumor core lower than 0.76 in the testing set. Hence, we decided to use the whole tumor contours for feature extraction.

The handcrafted signature achieved the C-index of 0.64, lower than the DL-based signature. However, the result of the paired t-test suggested no significant difference in the C-index. We also trained a fused Cox regression model to integrate the two signatures. But the fused signature achieved a slightly lower C-index (0.66) than the DL-based signature. The AUCs of both signatures evaluated at one-year OS were higher than 0.73. Our results showed that the DL-based signature, unlike the handcrafted signature, resulted in prognostically distinct groups using either X-tile generated or median threshold. Overall, the DL-based signature had a better association with GBM OS in terms of higher C-index and significant patient stratification.

Recently, a number of studies have proposed different 3D CNN architectures for achieving accurate GBM segmentation[24–26]. Accurate and precise tumor segmentation is essential for diagnosis, disease monitoring, and tumor characterization. The main goal of this work was to study the prognostic power of the radiomic features extracted from the automatically generated tumor contours. Other automatic segmentation methods can be integrated into the proposed workflow but were not explored and out of the scope of the paper. Potential future work includes selecting the best segmentation model and investigating whether more accurate automatically generated contours may result in a better survival prediction model.

Our study has several limitations. First, although we used an independent testing set for model evaluation, the number of training and testing patients is limited. This may result in suboptimal prediction performance and large evaluation biases. Larger datasets are required to further compare the two feature sets. Second, we only investigated the transfer learning method for extracting DL-based radiomic features due to the limited patient size. A CNN trained from scratch or using the fine-tuning technique for survival prediction could directly learn useful features from MR images. However, it may be easily overfitted and possibly require more patient data to achieve robust performance. Other methods like training an autoencoder for feature extraction would also be valuable to explore. Third, the information provided by the MR images may be limited and not powerful enough for achieving more accurate models. Future work could be done to include genomic features and investigate whether the combination of genomic and radiomic features could improve prediction performance.

## 5. Conclusion

We proposed an automatic DL-based workflow for GBM survival prediction using four pre-operative MR images. The proposed 3D CNN could generate accurate tumor contours. Our results showed that the DL-based radiomic features, extracted using the tumor contours automatically generated by the 3D CNN, achieved higher C-index than the handcrafted features. The constructed DL-based signature resulted in significant patient stratification. Our workflow demonstrated the potential of improving patient stratification and survival prediction in GBM patients.

**Acknowledgment**

This research was partially funded by Varian Medical Systems, Inc.

**Disclosure of Conflicts of Interest**

The authors have no relevant conflicts of interest to disclose.